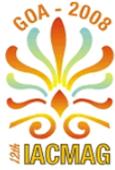

# Numerical upscaling of the permeability of a randomly cracked porous medium

S. Ghabezloo
*Université Paris-Est, UR Navier, CERMES-ENPC, Paris, France*

A. Pouya
*Université Paris-Est, LCPC, MSRGI, Paris, France*

ABSTRACT: The equivalent permeability of a randomly cracked porous material is studied using a finite element program in which a four-nodes zero-thickness element is implemented for modelling the cracks. The numerical simulations are performed for geometries with different cracks densities and for different values of matrix permeability and cracks conductivity, but the cracks length are taken equal to one. The method used for determination of the equivalent permeability resulted in a perfectly symmetric equivalent permeability tensor for each case. Based on the obtained results a simple relation is presented for the equivalent permeability of a randomly cracked porous material as a function of the matrix permeability and the cracks density and conductivity. This relation is then generalized for the cracks of any length using a linear transformation.

## 1 Introduction

Evaluation of the equivalent permeability of a cracked porous material has a great interest in geotechnical and petroleum engineering. In such material, the non-cracked matrix is permeable and the presence of the cracks, which are normally more permeable than the matrix, creates the preferential paths for water flow. When the cracks density is below the percolation threshold, the equivalent permeability of the cracked porous material is a function of the matrix permeability, cracks density and transmissivity and the statistical distribution of length and orientation of the cracks. The problem of equivalent permeability of a cracked porous material has been intensively investigated using the theoretical upscaling methods (Sánchez-Vila *et al.* 1995, Goméz-Hernández and Wen 1996, Renard and de Marsily 1997, Noetinger 2000, Dormieux and Kondo, 2002). Recently Pouya and Ghabezloo (2008) have presented a solution for the flow around a single crack in porous material and used this solution for upscaling the permeability of micro-cracked materials. But this approach, as well as other approaches using at best the self-consistent scheme for permeability upscaling, remains valid only for weak crack densities. In fact, these semi-analytical approaches neglect the crack interaction or represent it in a too simplified way. The upscaling of permeability for higher density of cracks requires adequate numerical modelling. In these methods the macroscopic permeability is deduced from the relations between the average velocity and pressure gradient in a Representative Elementary Volume. A review of recent works on numerical determination of the effective permeability of fractured rock masses gives witness, however, of difficulties in determining rigorously these average values for locally discontinuous materials. As a consequence, the compliance tensor obtained for the domain is found to be not symmetric and this is attributed wrongly to the finite-size of the domain. Pouya and Courtois (2002) and Pouya (2005) proposed a rigorous method for determining the mean flux and the mean pressure gradient from pressure and flux values on the boundary of the domain and showed that the equivalent permeability tensor obtained in this way is symmetric and positive-definite. In this work a numerical method based on Finite Elements is presented for a rigorous calculation of the effective permeability of a micro-cracked porous material. The analysis results in a law of equivalent permeability as a function of matrix permeability and cracks conductivity, length and density.

**Notation**: In what follows, light-face (Greek or Latin) letters denote scalars; underlined letters designate vectors and bold-face letters, second-order tensors. The convention of summation on repeated indices is used implicitly. The inner product of two vectors is labeled as $\underline{a}.\underline{b} = a_i b_i$ and the operation of a second-order tensor **a** on a vector $\underline{n}$ is labelled as $\mathbf{a}.\underline{n}$, $(\mathbf{a}.\underline{n})_i = a_{ij} n_j$.

## 2 Governing equations

A homogeneous bloc $\Omega$ containing a family of straight cracks $\Gamma$ is considered (Figure (1)). Fluid velocity $\underline{v}(\underline{x})$ in the matrix is given by the Dracy's law:

$$\forall \underline{x} \in \Omega - \Gamma \qquad \underline{v}(\underline{x}) = -\mathbf{k} \cdot \nabla p(\underline{x}) \qquad (1)$$

where $\mathbf{k}$ is the matrix permeability and $p(\underline{x})$ the pressure field. The *flux* (discharge) in the crack is given by a Poisieul type law, function of the pressure gradient along the crack:

$$\forall \underline{x} \in \Gamma \qquad q(s) = -c\, \partial p / \partial s \qquad (2)$$

Where $s$ is the abscise along the crack and $c$ is the crack's conductivity. Mass conservation in the matrix reads:

$$\forall \underline{x} \in \Omega - \Gamma \qquad \nabla \cdot \underline{v}(\underline{x}) = 0 \qquad (3)$$

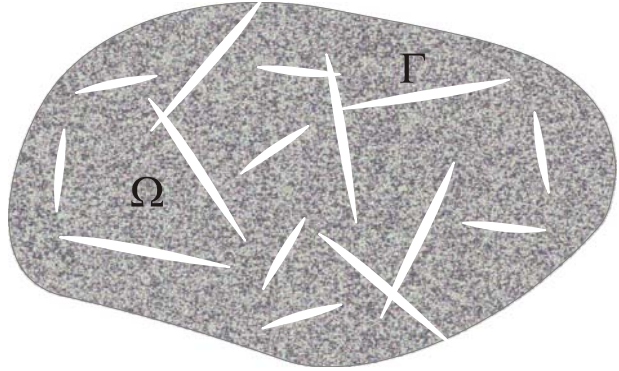

Figure 1. Porous matrix containing cracks

The crack-matrix mass exchange law reads, on running points on cracks (Pouya and Ghabezloo, 2008):

$$[\![\underline{v}(\underline{x})]\!] \cdot \underline{n} + \partial q / \partial s = 0 \qquad (4)$$

where $\underline{x}$ is the point in $\Omega$ corresponding to the abscise $s$ on the crack, $\underline{n}$ is the normal unit vector to the crack line and $[\![\cdot]\!]$ designates the discontinuity or jump across the crack.

2.1 Conductivity tensors for a finite size domain

A *linear pressure condition* $p(\underline{x}) = \underline{A} \cdot \underline{x}$ is prescribed on the boundary $\partial \Omega$, generating a velocity field $\underline{v}(\underline{x})$ in $\Omega$. The average velocity $\underline{V}$ is determined numerically by:

$$\underline{V} = \frac{1}{\Omega}\left[\int_\Omega \underline{v}\, d\Omega + \sum_i \int_{\Gamma^{(i)}} q\underline{t}\, ds\right] \qquad (5)$$

The linearity of equations implies that $\underline{V}$ is a linear function of $\underline{A}$. From $\underline{V}$ components obtained for two different directions of $\underline{A}$ we deduce the coefficients of the conductivity tensor $\mathbf{K}$ defined by $\underline{V} = -\mathbf{K} \cdot \underline{A}$. Pouya and Courtois (2002) and Pouya (2005) have shown, for continuous material, that this equivalent permeability matrix $\mathbf{K}$ is symmetric and positive-definite. This property can be extended to cracked material in the same way that for mechanical properties as shown by Pouya and Chalhoub (2008).

## 3 Finite element modelling

In order to study the problem of equivalent permeability of a cracked porous material, a 2D finite element program is developed for solving the equations of fluid transfer in a double porosity material, presented in the preceding section. A four-node zero-thickness joint element is integrated into the program for the modelling of the cracks. The statistical distribution is limited to randomly oriented cracks. The equivalent permeability in this case is isotropic. A square shaped REV is considered and a family of random cracks with a given density is generated within the REV (Figure 2). The cracks density $\rho$ is defined by:

$$\rho = \frac{1}{\Omega} N \left(l/2\right)^2 \tag{6}$$

where $N$ is the number of cracks, $l$ is the cracks length and $\Omega$ is the area of REV. The length of the cracks is taken equal to one and the size of the REV is defined with some preliminary calculations equal to ten. After the generation of the cracks with random positions and orientations, the cracks that have a part outside the REV are truncated. Then the finite element mesh is generated and the nodes on the cracks are doubled to form four-node zero-thickness joint elements. In order to optimize the calculation time, the nodes and elements are renumbered after the creation of joint elements. An example of geometry with cracks density $\rho = 1$ is presented in Figure (2).

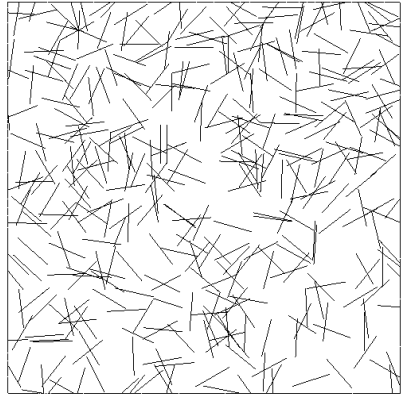

Figure 2. Example of geometry with $\rho$=1

The linear pressure boundary condition $p(\underline{x}) = \underline{A} \cdot \underline{x}$ is applied to the model. Using the velocity vectors of the matrix elements and the discharge of the joint elements in equation (5) the average velocity vector $\underline{V}$ is calculated. By solving the problem for two different vectors $\underline{A}_1$ and $\underline{A}_2$ two velocity vectors $\underline{V}_1$ and $\underline{V}_2$ are obtained. Knowing that $\underline{V} = -\mathbf{K} \cdot \underline{A}$, the four elements of $\mathbf{K}$ can be found by solving a system of two linear equations. The method of determination of equivalent permeability tensor presented here above (Pouya and Courtois, 2002; Pouya, 2005) is resulted in an equivalent permeability tensor $\mathbf{K}$ which is perfectly symmetric. As an example, for a model with $\rho = 1.0$ (Figure (2)), $k = 1 \times 10^{-8}$ m/sec, $c = 1 \times 10^{-6}$ m$^2$/sec and thus $c/k = 100$ the calculated equivalent permeability tensor is:

$$\mathbf{K} = \begin{bmatrix} 1.3223 \times 10^{-7} & -8.1411 \times 10^{-9} \\ -8.1411 \times 10^{-9} & 1.2589 \times 10^{-7} \end{bmatrix} \tag{7}$$

Due to the random distribution of the position and orientation of the cracks and also the isotropic permeability of the porous matrix ($k_{xx} = k_{yy} = k$, $k_{xy} = k_{yx} = 0$), the elements $K_{xx}$ and $K_{yy}$ of the equivalent permeability matrix $\mathbf{K}$ have very close values. The average of these two values is considered as the equivalent permeability of the cracked porous material. For a more general case we can write:

$$K^{eq} = \sqrt{|\mathbf{K}|} \tag{8}$$

where $|\mathbf{K}|$ is the determinant of the second-order matrix $\mathbf{K}$. Thus for the equivalent permeability tensor presented in equation (7), the ratio of the equivalent permeability of the cracked medium to the matrix permeability, $K^{eq}/k$ is obtained equal to 12.91. For each finite element model with a given cracks density, the equivalent permeability tensor is determined for different values of $c/k$ ratio and the results are presented in the following section.

## 4  Modelling results

In order to study the effect of the cracks density and conductivity on the equivalent permeability of the cracked porous material, 18 different models with cracks densities between 0.2 and 1.2 are prepared. For each model the equivalent permeability is calculated for the ratio of the cracks conductivity over matrix permeability, $c/k$, equal to 1, 5, 10, 50 and 100 which make a total of 90 data points. Figure (3) shows the ratio of the equivalent permeability to the matrix permeability, $K^{eq}/k$, as a function of the cracks density $\rho$ for different values of cracks conductivity to matrix permeability ratio, $c/k$. We can see the increase of the $K^{eq}/k$ ratio with the cracks density and conductivity. To obtain a unique expression for the equivalent permeability as a function of the cracks density and conductivity, Figure (4) presents the expression $\mathrm{Ln}(K^{eq}/k)/\rho$ as a function of the $c/k$ ratio, shown in a logarithmic scale, from which the following expression is obtained for the equivalent permeability:

$$\frac{1}{\rho}\mathrm{Ln}(K^{eq}/k) = 0.38\,\mathrm{Ln}(c/k) + 0.92 \tag{9}$$

Note that this relation is valid for high conductivity cracks ($c/k > 1$) and is not valid for the limit $c \to 0$.

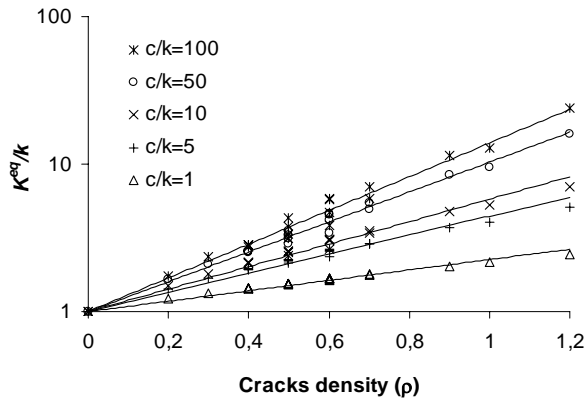
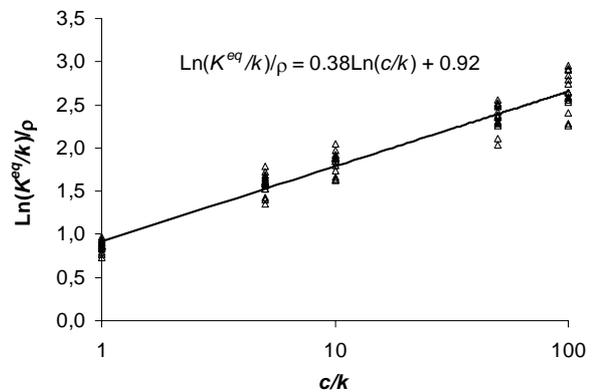

Figure 3. Increase of the equivalent permeability with cracks density for different values of *c/k*

Figure 4. Increase of the equivalent permeability, normalized with matrix permeability and cracks density, with *c/k* ratio

Equation (9) can be re-written in the following form:

$$\frac{K^{eq}}{k} = \left[2.51\left(\frac{c}{k}\right)^{0.38}\right]^{\rho} \tag{10}$$

Equation (10) gives the equivalent permeability of a randomly cracked porous material as a function of the matrix permeability and cracks density and conductivity, but is restricted to the case in which the cracks length is equal to one. This equation will be generalized in the following section for the cracks of every length using a theoretical transformation. The results then will be validated using some more numerical calculations.

## 5 Linear transformation

Consider the problem presented in Figure (5). Here we consider an isotropic permeability tensor for the porous matrix. Considering the isotropic distribution of the cracks, the resulted equivalent permeability tensor will also be isotropic. So we replace the second-order tensors $\mathbf{k}$ and $\mathbf{K}$ respectively with the scalar values $k$ and $K^{eq}$. As described before the solution of the problem to find the fields $\underline{v}(\underline{x})$ and $p(\underline{x})$ is given by the following equations:

$$\begin{aligned}
p(\underline{x}) &= \underline{A} \cdot \underline{x} & \forall \underline{x} \in \partial\Omega \\
\underline{v}(\underline{x}) &= -k\nabla P(\underline{x}) & \forall \underline{x} \in \Omega - \Gamma \\
q(s) &= -c\,\partial p/\partial s & \forall \underline{x} \in \Gamma \\
\nabla \cdot \underline{v}(\underline{x}) &= 0 & \forall \underline{x} \in \Omega - \Gamma \\
[\![\underline{v}(\underline{x})]\!] \cdot \underline{n} + \partial q/\partial s &= 0 & \forall \underline{x} \in \Gamma
\end{aligned} \tag{11}$$

Now let introduce a linear transformation of the coordinate system such that:

$$\underline{\tilde{x}} = \lambda^{-1}\underline{x} \tag{12}$$

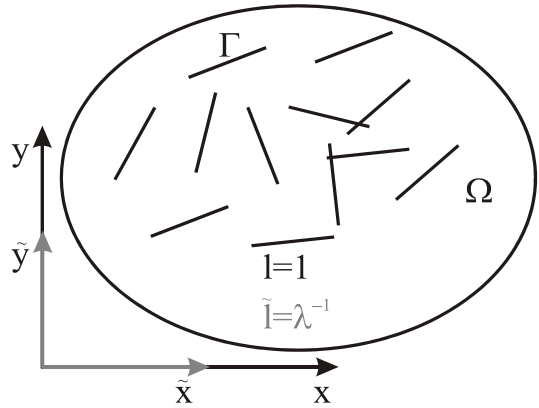

Figure 5. Linear transformation

This transformation changes the domain $\Omega$ to a domain $\tilde{\Omega}$ and the cracks $\Gamma$ with length equal to one to the cracks $\tilde{\Gamma}$ with length equal to $\lambda^{-1}$. Now we define a new solution field such that:

$$\underline{\tilde{v}}(\underline{\tilde{x}}) = \underline{v}(\underline{x}) \quad, \quad \tilde{p}(\underline{\tilde{x}}) = p(\underline{x}) \tag{13}$$

For this new solution we have:

$$\tilde{\nabla}\tilde{p}(\underline{\tilde{x}}) = \frac{\partial \tilde{p}(\underline{\tilde{x}})}{\partial \underline{\tilde{x}}_i} = \lambda \frac{\partial p(\underline{x})}{\partial \underline{x}_i} = \lambda \nabla p(\underline{x}) \tag{14}$$

and so:

$$\underline{\tilde{v}}(\underline{\tilde{x}}) = -\tilde{k}\tilde{\nabla}\tilde{p}(\underline{\tilde{x}}) = -\tilde{k}\,\lambda\nabla p(\underline{x}) \tag{15}$$

From equation (13) we know that $\underline{\tilde{v}}(\underline{\tilde{x}}) = \underline{v}(\underline{x})$ and from equation (11) we have $\underline{v}(\underline{x}) = -k\nabla p(\underline{x})$. So the following relation between the matrix permeabilities is found:

$$\tilde{k} = \lambda^{-1}k \tag{16}$$

From equation (13) for the points on the cracks we have $[\![\tilde{v}(\tilde{x})]\!] \cdot \tilde{n} = [\![v(x)]\!] \cdot n$ and so to satisfy equation (11) for the transformed problem we should have $\partial \tilde{q}/\partial \tilde{s} = \partial q/\partial s$. Knowing that $s = \lambda \tilde{s}$ we obtain $\partial \tilde{q}/\partial \tilde{s} = \lambda \, \partial \tilde{q}/\partial s$ and so we find:

$$\tilde{q} = \lambda^{-1} q \tag{17}$$

For the points on the cracks from equation (11) we have $q = -c \, \partial p/\partial s$. Knowing that $s = \lambda \tilde{s}$ and $\tilde{p}(\tilde{x}) = p(x)$ we find $\partial p/\partial s = \lambda^{-1} \partial \tilde{p}/\partial \tilde{s}$ and so $q = -c \lambda^{-1} \partial \tilde{p}/\partial \tilde{s}$. Replacing this in equation (17) we find the following relation between the crack conductivities:

$$\tilde{c} = \lambda^{-2} c \tag{18}$$

From the definition given in equation (6) and knowing that $l = \lambda^{-1}$ and $\tilde{\Omega} = \lambda^{-2} \Omega$ the cracks density for the transformed problem is written as the following:

$$\tilde{\rho} = \frac{N l^2}{4 \tilde{\Omega}} = \frac{N \lambda^{-2}}{4 \lambda^{-2} \Omega} = \frac{N}{4\Omega} = \rho \tag{19}$$

Using equations (12) and (13), the mass conservation equation for $\tilde{\Omega}$ is written in the following form:

$$\tilde{\nabla} \cdot \tilde{v}(\tilde{x}) = \frac{\partial \tilde{v}(\tilde{x}_i)}{\partial \tilde{x}_i} = \lambda \frac{\partial v(x_i)}{\partial x_i} = \lambda \nabla \cdot v(x) = 0 \tag{20}$$

So we showed the solution fields $v(x)$ and $p(x)$ corresponding to a cracked porous material $\Omega$ with the matrix permeability $k$, cracks conductivity $c$, length equal to one and density $\rho$ are equivalent to the solution fields $\tilde{v}(\tilde{x})$ and $\tilde{p}(\tilde{x})$ of another problem $\tilde{\Omega}$ with the matrix permeability equal to $\lambda^{-1} k$, cracks conductivity equal to $\lambda^{-2} c$, length equal to $\lambda^{-1}$ and the same density $\rho$.

The average velocity vector for the problem $\tilde{\Omega}$ can be obtained using equation (5):

$$\tilde{V} = \frac{1}{\tilde{\Omega}} \left[ \int_\Omega \tilde{v} d\tilde{\Omega} + \sum_i \int_{\tilde{\Gamma}^{(i)}} \tilde{q} \tilde{t} d\tilde{s} \right] = \frac{1}{\lambda^{-2}\Omega} \left[ \int_\Omega v \lambda^{-2} d\Omega + \sum_i \int_{\Gamma^{(i)}} \lambda^{-1} q t \lambda^{-1} ds \right] = V \tag{21}$$

For the boundary $\partial \Omega$ we have $p(x) = A \cdot x$. Knowing that $\tilde{p}(\tilde{x}) = p(x)$ and using equation (12) we find:

$$\tilde{A} = \lambda A \tag{22}$$

Using equations (21) and (22) and knowing that $V = -K^{eq} A$ we can write $\tilde{K}^{eq} \tilde{A} = K^{eq} A = K^{eq} \lambda^{-1} \tilde{A}$ and so we obtain the following relation between the equivalent permeability tensors:

$$\tilde{K}^{eq} = \lambda^{-1} K^{eq} \tag{23}$$

Let us notice that $\tilde{K}^{eq}$ is the equivalent permeability for the case of local parameters $\tilde{k}$, $\tilde{c}$, $\tilde{\rho}$ and $\tilde{l}$. This means that we can write $\tilde{K}^{eq}(\tilde{k},\tilde{c},\tilde{\rho},\tilde{l}) = K^{eq}(k,c,\rho,l)$. So equation (23) can be written in the following from:

$$K^{eq}(\lambda^{-1}k, \lambda^{-2}c, \rho, l) = \lambda^{-1} K^{eq}(k, c, \rho, 1) \tag{24}$$

We can re-write equation (24) for a new set of variables by considering $k' = \lambda^{-1} k$ and $c' = \lambda^{-2} c$ and knowing

$l = \lambda^{-1}$:

$$K^{eq}(k',c',\rho,l) = lK^{eq}(l^{-1}k',l^{-2}c',\rho,1) \qquad (25)$$

Now we can replace the variable $k'$ and $c'$ respectively by $k$ and $c$ to have:

$$K^{eq}(k,c,\rho,l) = lK^{eq}(l^{-1}k,l^{-2}c,\rho,1) \qquad (26)$$

From equation (11) it is evident that if we multiply the matrix permeability $k$ and the cracks conductivity $c$ by a coefficient $\alpha$, the resulted equivalent permeability will be multiplied by $\alpha$. So we have:

$$K^{eq}(k,c,\rho,l) = \alpha^{-1}K^{eq}(\alpha k,\alpha c,\rho,l) \qquad (27)$$

By taking $\alpha = k^{-1}$ in equation (27) we find:

$$K^{eq}(k,c,\rho,l) = kK^{eq}(1,k^{-1}c,\rho,l) \qquad (28)$$

Once again by taking $\alpha = k^{-1}l$ we obtain:

$$K^{eq}(l^{-1}k,l^{-2}c,\rho,1) = kl^{-1}K^{eq}(lk^{-1}l^{-1}k,lk^{-1}l^{-2}c,\rho,1) = kl^{-1}K^{eq}(1,k^{-1}l^{-1}c,\rho,1) \qquad (29)$$

Replacing equation (28) and (29) in equation (26) we obtain:

$$K^{eq}(1,k^{-1}c,\rho,l) = K^{eq}(1,k^{-1}l^{-1}c,\rho,1) \qquad (30)$$

We can simply re-write equation (30) in the following form:

$$K^{eq}\left(\frac{c}{k},\rho,l\right) = K^{eq}\left(\frac{c}{kl},\rho\right) \qquad (31)$$

Using equation (31) the expression of equation (10) can be generalized for the case of cracks of length $l$ in the following form:

$$\frac{K^{eq}}{k} = \left[2.51\left(\frac{c}{kl}\right)^{0.38}\right]^{\rho} \qquad (32)$$

The generalized expression presented in equation (32) can be verified using some numerical simulations. For doing this two different cases are compared: The first is with $l=1$ and $c/k=10$ and the second one is with $l=10$ and $c/k=100$ and so for both cases $c/kl=10$. Figure (6) shows the increase of the $K^{eq}/k$ ratio with the cracks density $\rho$ for both cases. The perfect superposition of the results of the two studied cases verifies the presented transformation (equation (32)).

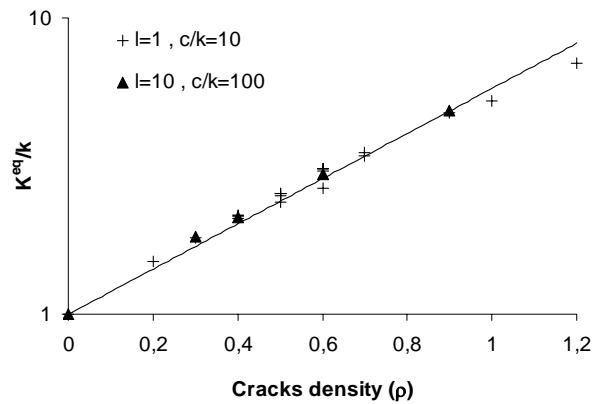

Figure 6. Verification of the linear transformation

## 6  Conclusion

The equivalent permeability of a cracked porous material is studied numerically using a finite element program in which the cracks of equal length are modelled using zero-thickness four-nodes elements. A random distribution is considered for positions and orientations of the cracks. A linear pressure boundary condition is applied to the model and the equivalent permeability is calculated using the method proposed by Pouya and Courtois (2002) and Pouya (2005). This method resulted in a perfectly symmetric equivalent permeability tensor. Several numerical simulations are performed for geometries with different cracks density and for different ratios of the cracks conductivity over the matrix permeability. For each case the equivalent permeability tensor is calculated for the cracks with the length equal to one. Based on the obtained results a simple equation is presented for the equivalent permeability of a randomly cracked porous material as a function of the matrix permeability and the cracks density and conductivity for the cracks length equal to one. This relation is then generalized for the cracks of any length using a linear transformation. The presented transformation is then verified by some more numerical simulations.

## 7  References

Dormieux L., Kondo D. 2004. Approche micromécanique du couplage perméabilité–endommagement, *C.R. Mecanique* 332, 135-140

Goméz-Hernández J.J., Wen X.H. 1996. Upscaling hydraulic conductivities in heterogeneous material: an overview, *J. Hydrol.* 183 ix–xxxii.

Noetinger B., 2000. Computing the effective permeability of lognormal permeability fields using renormalization methods, *C. R. Acad. Sci. Paris*, Ser. IIa 331, 353–357.

Pouya A., Ghabezloo S. 2008. Flow around a crack in a porous matrix and related problems, *Transport in Porous Media*, (Submitted).

Pouya A., Courtois A. 2002. Definition of the permeability of fractured rock masses by homogenization methods, *C.R. Géosciences,* 334 (13), 975-979.

Pouya A. 2005. Equivalent permeability tensors of a finite heterogeneous block, *C.R. Géosciences,* 337 (6), 581-588.

Pouya A. Chalhoub M. 2008. Numerical homogenization of elastic behavior of fractured rock masses and micro-cracked materials by FEM, *The 12th international conference of International Association for Computer Methods and Advances in Geomechanics (IACMAG), 1-6 octobre 2008, Goa, India*

Renard P., de Marsily G. 1997. Calculating equivalent permeability: a review, *Adv. Water Resour.* 20 (5–6) 253–278.

Sánchez-Vila X., Girardi G.P., Carrera J. 1995. A synthesis of approaches to upscaling of hydraulic conductivities, *Water Resour. Res.* 31 (4) 867–882.